\documentclass[epj
]{svjour}
\usepackage{graphicx}
\newcommand\beq{\begin{equation}}
\newcommand\eeq{\end{equation}}
\begin{document}
\title{How does flow in a pipe become turbulent?}
\author{Bruno Eckhardt\inst{1} \and Tobias M. Schneider\inst{1}
}                     
%
%
\institute{Fachbereich Physik, Philipps-Universit\"at Marburg, 35043 Marburg}
\date{Received: date / Revised version: date}
%
\abstract{The transition to turbulence in pipe flow does not follow
  the scenario familiar from Rayleigh-Benard or Taylor-Couette flow
  since the laminar profile is stable against infinitesimal
  perturbations for all Reynolds numbers. Moreover, even when the flow
  speed is high enough and the perturbation sufficiently strong such
  that turbulent flow is established, it can return to the laminar
  state without any indication of the imminent decay. In this parameter
range, the lifetimes of perturbations show
 a sensitive dependence on initial
  conditions and an exponential distribution. The turbulence seems to
  be supported by three-dimensional travelling waves
  which appear transiently in the flow field. The boundary between laminar and
  turbulent dynamics is formed by the stable manifold of an invariant
  chaotic state. We will also discuss the relation between
  observations in short, periodically continued domains, and the
  dynamics in fully extended puffs.
\PACS{
{47.27.Cn}{ Transition to turbulence} \and
{47.27.nf}{ Flows in pipes and nozzles} \and
{47.10.Fg}{Dynamical systems methods} 
}
} 
\maketitle
\section{Introduction}
\label{intro}
Dynamical system theory and nonlinear dynamics has figured prominently
in understanding the transition to turbulence in systems with linear
instabilities of the laminar profile \cite{kosc93}.  The various
bifurcations and transitions in fluids heated from below
(Rayleigh-Benard) and in the flow between independently rotating
cylinders (Taylor-Couette) have been studied in considerable details
and have helped to advance both our understanding of the various
transition scenarios and of the properties of dynamical systems in
general.  Pressure driven flow in a pipe or the flow between moving parallel
plates (plane Couette flow) behave differently and are less well
understood: the linear stability of the laminar profile prevents the
immediate application of the usual bifurcation and transition
scenarios \cite{boberg88,grossmann00,Ker05,Eck07b}.  So how does the turbulence
observed in experiments come about?

Evidence that has emerged in recent years 
suggests that dynamical systems also provide a role model for the transition in
linearly stable situations \cite{bruno02,Eck07b,aberdeen}. 
The application of these ideas to the typically fairly high-dimensional 
situation of pipe flow also raises
interesting questions about dynamical systems, so that one can again
expect a fruitful interplay between dynamical systems theory and the
fluid dynamics of the transition to turbulence. The focus of this
contribution will be to survey recent progress in our understanding of
pipe flow.  
We will go beyond previous studies \cite{bristol_proc,Ker05,Eck07b} 
by including steps
towards a characterization of the spatial variability of the system.

The main features of turbulence transition in a pipe were documented
long ago by Osborne Reynolds \cite{Rey83,Rey83a}. He noticed the
possibility to delay the transition by suppressing perturbations in
the inflow region, the intermittent character of the transition, and
the presence of vortices in the turbulent regions. He noticed that the
flow can be characterized by a dimensionless combination of mean
downstream velocity $U$, radius of the pipe $R$ and kinematic
viscosity $\nu$, nowadays known as the Reynolds number $Re=2U R/\nu$.
The intermittent character of the transition and the absence of a
linear instability of the laminar profile make the determination of a
`critical Reynolds number' above which turbulence can be observed a
tricky business, as attested by the huge variability of critical
numbers that can be found in the literature.  Reynolds arrives at
about 1800, Prandtl \cite{Prandtl} speculates that it should be 
above 1000, Mullin and Darbyshire \cite{darbyshire95} 
find long lived turbulent trajectories above about
1650, and most reference books and textbooks quote numbers in the
range of 2000 to 3000. The operational definition would be that the
critical Reynolds number is defined such that sufficiently strong
perturbations can induce turbulence if the critical value is exceeded.
But even with that definition there remain problems connected 
with the transience of the turbulent state, as discussed
in section 5.

An interesting feature of turbulence in pipe flow is that below
Reynolds numbers of about 2700, it does not extend throughout the pipe
but remains localized in turbulent puffs of about $60R$ length, 
\cite{wygnanski73,wygnanski75}. 
While some progress has been made in describing localized structures anf
frontal dynamics in various pattern forming systems \cite{crosshohen},
the case of localized turbulence is more complicated because of the 
dynamic nature of the localized state \cite{Sch01}. Some progress has
recently been made for the case of localized turbulent stripes in 
Taylor-Couette and plane Couette flow, \cite{prigent,barkley}.

For our presentation below we will draw on some experimental observations,
and on numerical simulations of the full Navier-Stokes equation in
domains of length 100R (for the spatially extended puffs) and 10R 
(for the statistical studies in smaller domains), where $R$ is the radius.
The numerical code has been verified by reproducing the spectrum of the
linearized equation, and the turbulent properties at somewhat higher Reynolds.
The coherent structures have also been confirmed by independent studies.

The outline of the paper is as follows. In section 2 we discuss the
dynamics of a localized turbulent puff and present evidence that the
interior dynamics can be captured by studies in smaller domains. In
section 3 we discuss coherent states and their bifurcations in domains
of length 10R. In secion 4 we present lifetime studies in the
small domain and relate them to experiment and other investigations.
In section 5 we discuss the boundary between the laminar and turbulent
regions in state space and the evidence for the presence of an edge of
chaos as a generalization to the usual basin boundaries. In section 6
we summarize and give a brief outlook how this relates to studies in
other systems.

\section{Puffs}

\begin{figure}
\begin{center}
\includegraphics[width=8.2cm,angle=0]{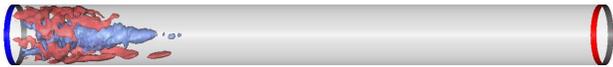}
\end{center}
\caption[]{
Three-dimensional flow inside a puff  at $Re = 1825$ moving from left to right. 
The snapshots are separated by time intervals of $t = 40 R/ u_{cl}$ and show 
isosurfaces of the downstream velocity relative to the laminar profile at levels  
$\pm 0.1 u_{cl}$ (negative in blue and positive in red). The plots are compressed by 
a factor of $5$ in the axial direction to show the whole computational domain of 
$L = 100 R$. 
{\bf Three more frames not included for size reasons.}
\label{puff}
}
\end{figure}

\begin{figure}
\begin{center}
\includegraphics[width=6.0cm,angle=0]{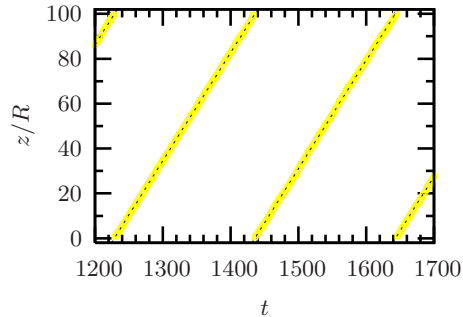}
\end{center}
\caption[]{
  Uniform motion of the puff along the pipe. The yellow line is the
  position of the center of turbulent energy in the in-plane velocity
  components, the dashed line a linear approximation to it. Shown are
  two passages through the pipe (which in the numerical simulations is periodically
  continued along the axis). The Reynolds number is
  $Re=1800$.
\label{motion}
}
\end{figure}

In the Reynolds number range up to about 2700, the transition to
turbulence in pipe flow shows a striking spatial localization
\cite{wygnanski73,wygnanski75}: the
turbulence is concentrated in regions of about $60$R length, which
move uniformly through the pipe
\cite{wygnanski73,wygnanski75,Mul05,Mul06,Pei06,Pei07,bristol_hof}.
In Fig.~\ref{puff} we present four snapshots of a puff and
in Fig.~\ref{motion} the axial position of its center, demonstrating
the very uniform speed with which it moves downstream.
The position of the puff is defined via the center of the turbulent
in-plane kinetic energy distribution. This criterion is very robust
as compared to one based on a jump in the downstream
centerline velocity (see \cite{Wil07a}), and gives almost
noise-free position information and very accurate translation velocities, 
as evident from Fig.~\ref{motion}, where no smoothing or linear 
interpolation was applied. 

The axial extension of the structure remains fairly constant and its
traveling velocity decreases form $0.965$ $U$ of the mean downstream
velocity at Re=1800 to $0.94$ $U$ at Re=1900. Thus, the turbulent 
patch is slightly slower than the bulk velocity of the fluid. 
Consequently, when viewed from a frame of reference comoving with the
puff, there is a net flow of fluid from the trailing to the leading edge 
of the puff.

\begin{figure}
\begin{center}
\includegraphics[width=8.0cm,angle=0]{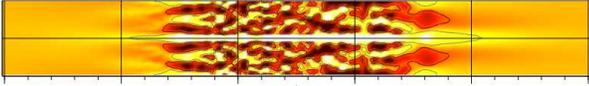}
\end{center}
\caption[]{Coherent structures in a turbulent puff at three different
	times for $Re=1800$. The abscissa runs from $z=10 R$ to $z=60 R$ (half 
	of the computational domain of $L=100 R$) and is translated so that
	the center of the turbulent intensity is kept at $z=L/3$. The ordinate 
	shows the angle $\phi$ from $-\pi$ to $\pi$ with the $0$ level indicated
	by the horizontal line. White shading indicates
  positive values, dark shading negative values of the correlation
  function $C(\phi,z)$. Note the rather disordered interior and the
  elongated coherent structures at the upstream and downstream
  edges of the puff.
  {\bf Two more frames not included because of size problems.}
\label{coherent}
}
\end{figure}

The velocity field inside the puff can be divided up into three
regions, a fairly homogeneous turbulent interior bracketed by the
upstream and downstream front regions. In order to highlight the different
characteristics of the velocity fields in these domains, we apply the
correlation function measures introduced in \cite{Sch07b} to extract
coherent structures.  Since in the situation of a localized puff the
system is not only inhomogeneous in wall-normal but also in axial
direction we only take out the mean velocity as obtained from
averages over the azimuthal direction, the only homogeneous direction left. 
Thus, the correlation functions are evaluated for the local deviations 
from the downstream speed as calculated along a ring at radius $r=0.81$:
\beq 
\tilde u_z(\phi,z)=u_z (r=0.81,\phi,z)-\langle
u_z (r=0.81,\phi,z) \rangle_\phi 
\eeq 
and 
\beq C(\phi,z)=\langle\tilde
u_z(\psi+\phi,z) \tilde u_z (\psi,z) \rangle_\psi
\label{indicator}
\eeq
Snapshots of this correlation function in Fig.~\ref{coherent} 
clearly show the three different regions. 
Intriguingly,
the upstream and downstream regions are very often dominated by 
pronounced coherent structures. On the upstream side, structures with
three and four streaks dominate. Thus, the center panel of 
Fig.~\ref{coherent} shows a typical flow configuration.
On the downstream side, the
structures are somewhat shorter and occur not quite as often
as on the upstream side. The dynamics of these coherent structures
and their significance for the transition to turbulence was studied 
extensively by Cas van Doorne in his PhD thesis, \cite{doorne_thesis}. 
The central region is spatially and temporally more disordered.

\begin{figure}
\begin{center}
\includegraphics[width=6.0cm,angle=0]{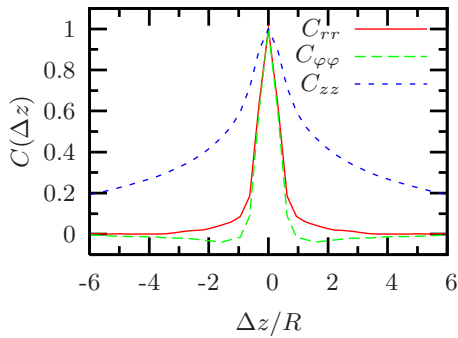}
\end{center}
\caption[]{
  Autocorrelation functions for the downstream ($C_{zz}$), azimuthal
  ($C_{\varphi \varphi}$) and radial ($C_{rr}$) velocities along the
  axis evaluated in the comoving frame of reference. The correlations are
  normalized to one for vanishing axial shift $\Delta z$. They  
are based on the deviation from the time and area 
  averaged mean profiles, resolved along the axis. The correlation functions
  are averaged over $3/4$ of the computational domain. They include the 
 entire puff, and thus also the elongated coherent structures at the up- and
  downstream boundaries.
  The time average is based on about 1000 independent velocity
  fields. The presented curves are for a radial position of $r=0.73$,
  but not much variation with $r$ was observed. 
\label{correlation}
}
\end{figure}

The disorder in the interior domain is also reflected in rapidly
decaying spatial auto-correlation functions of the three velocity
components, see Fig.~\ref{correlation}.  The downstream component
shows the longest axial correlation length due to the presence of
streaks. As measured by the level it reaches $1/$e, it extends over a 
width of about $2 R$.  
The correlation functions for the other components drop to $1/$e of
their maximal value over distances of less than one pipe radius. The
short axial correlation length suggests that useful
information about the interior dynamics of the flow can be obtained by
studying relatively short domains. Specifically, 
for a length of $10R$ the axial correlations in the downstream fluid 
are down to $30$ per cent, but the computational advantages
are enourmous and allow for detailed studies of deterministic \cite{Fai03}
and statistical analyses \cite{Sch07b}.  

\section{Local dynamics in short domains}
The correlation functions of the previous section suggest that it
should be possible to study the dynamics in short domains, 
thereby eliminiating much of the large-scale
spatial dynamics.  Nevertheless, the state space of the system remains
fairly high-dimensional, and easily reaches ${\cal O}(10^5)$ active dynamical 
degrees of freedom after the elimination of boundary conditions and 
incompressibility.

The coexistence of laminar and turbulent dynamics for the same
parameter values but different initial conditions in this system
suggests that the transition might be associated with a subcritical
bifurcation (compare \cite{ott93}). For increasing Reynolds number,
the usual scenario would then imply the appearance of new states in a
saddle-node bifurcation, followed by the destabilization of
the laminar profile in a collision with the saddle (as realized, for
instance, in TC flow in the narrow gap limit, \cite{Fai00}). 

\begin{figure}
\begin{center}
\includegraphics[width=3.cm,angle=0]{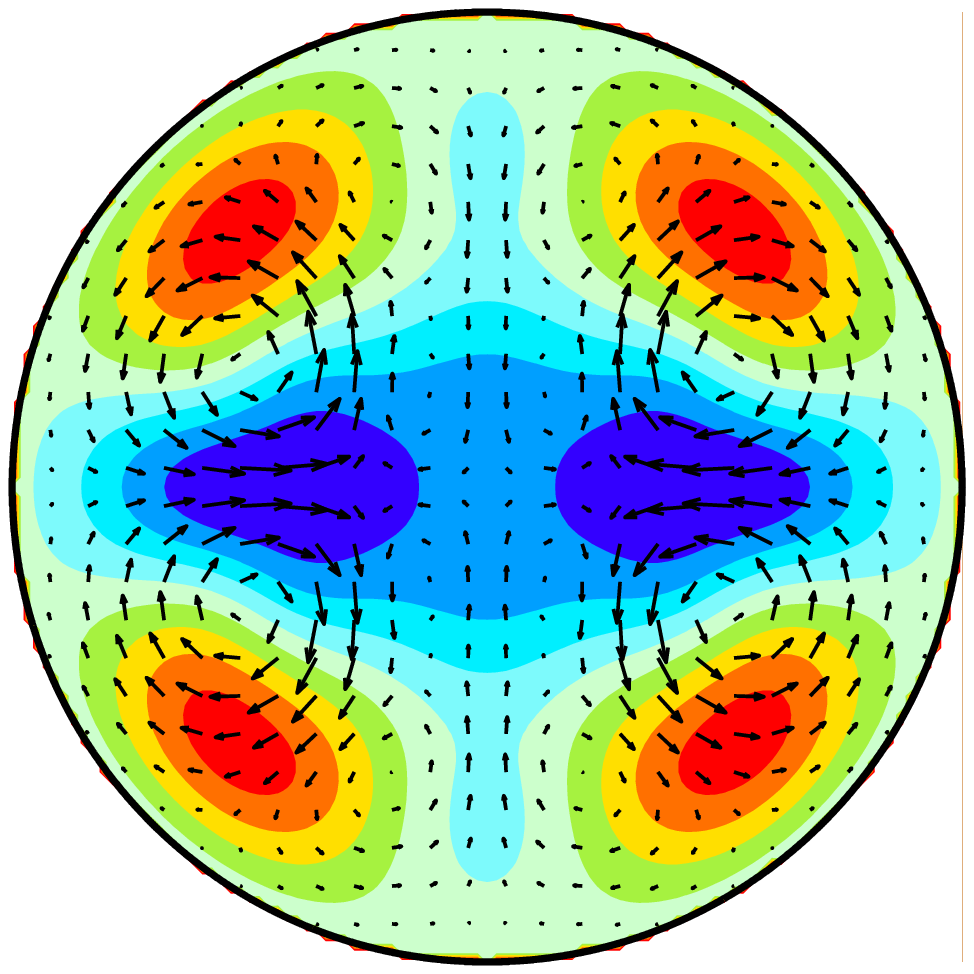}
\includegraphics[width=3.cm,angle=0]{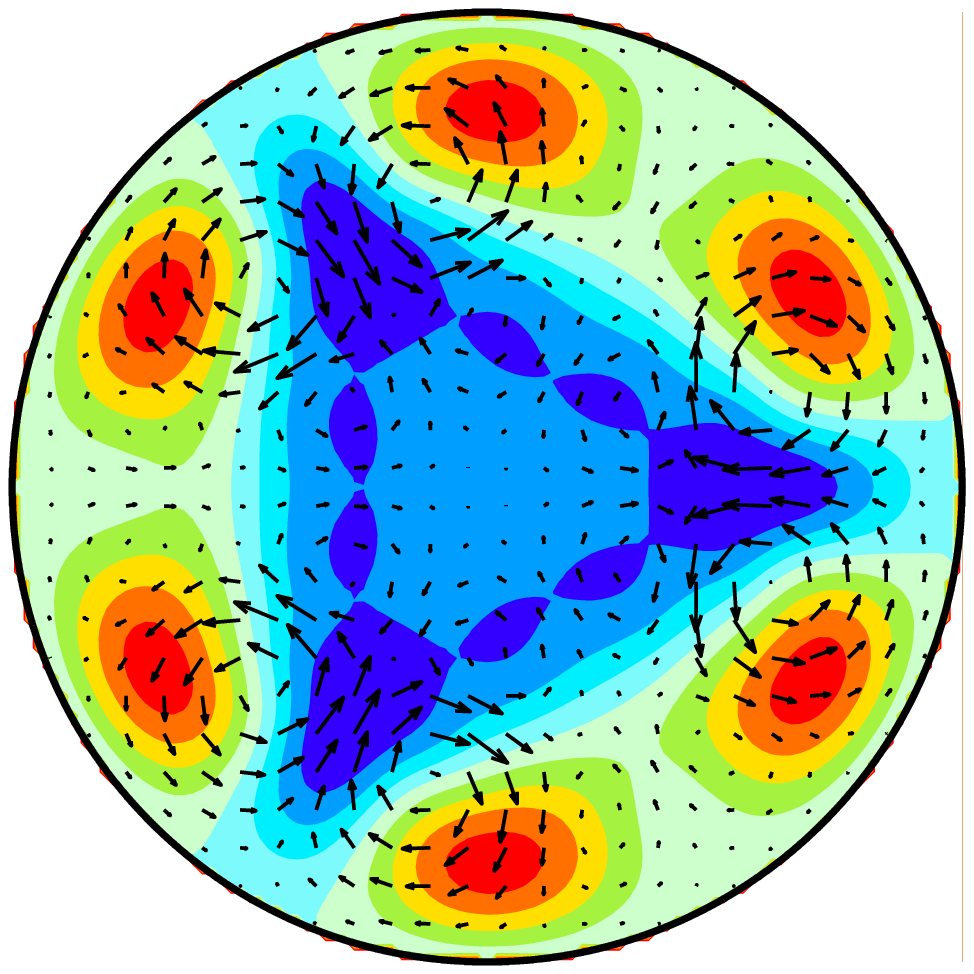}
\includegraphics[width=3.cm,angle=0]{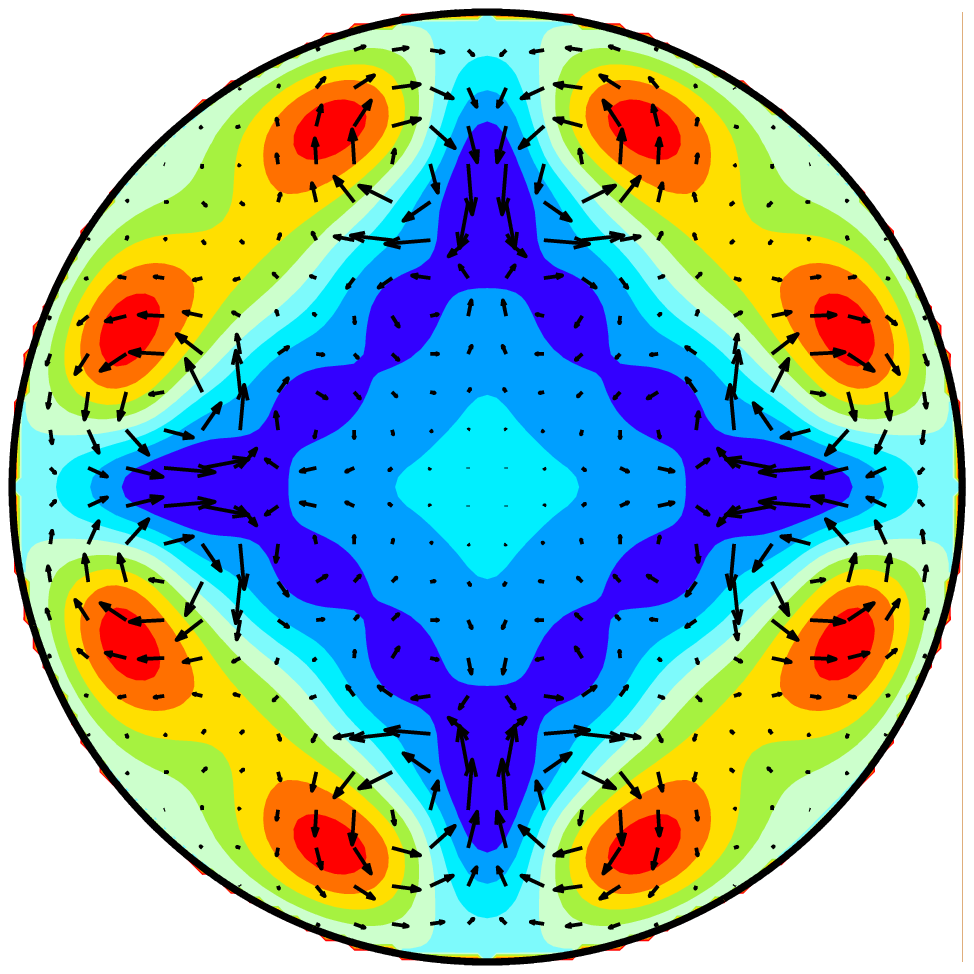}
\includegraphics[width=3.cm,angle=0]{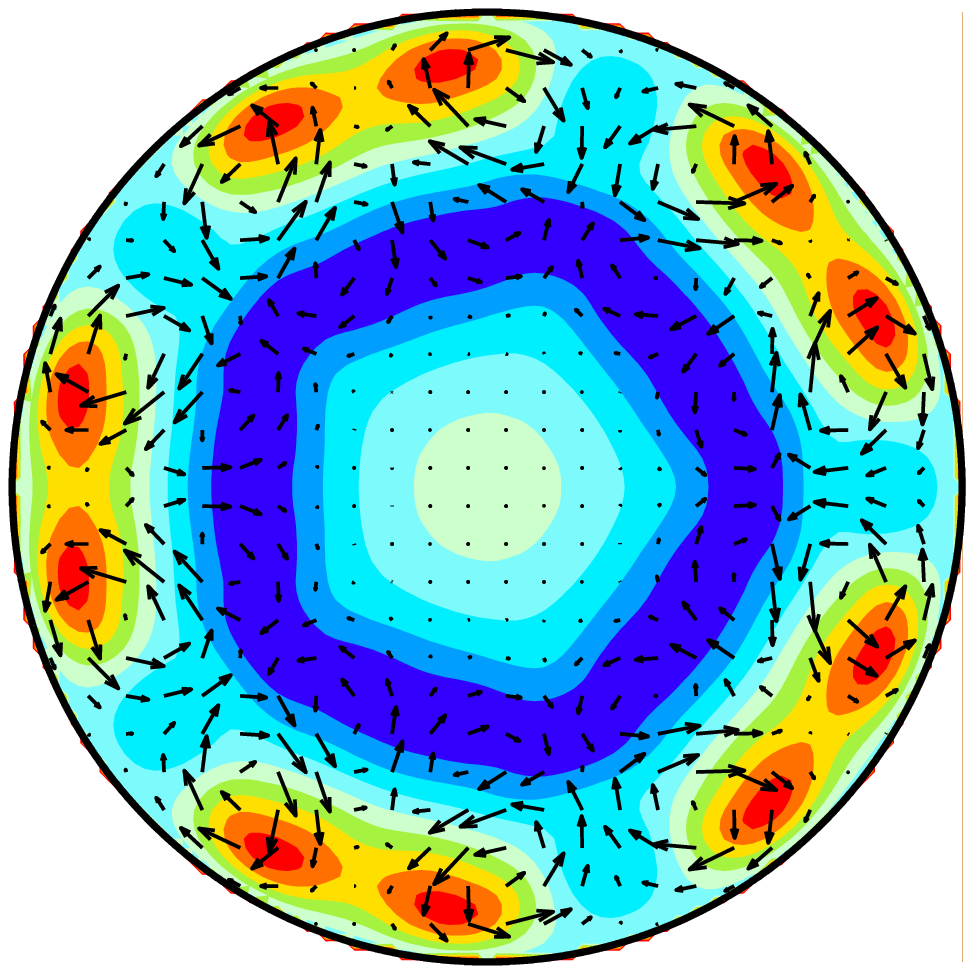}
\end{center}
\caption[]{Coherent states for pipe flow. Shown are velocity profiles averaged in the 
  downstream direction. The deviation from the laminar Hagen-Poseuille
  profile is indicated by colors. Negative velocities are shown in
  blue, positive areas in red. In-plane velocity components are
  indicated by vectors. The regular arrangement of high- and
  low-speed streaks reflects different discrete rotational symmetry
  classes.
\label{coherent_states}
}
\end{figure}

\begin{figure}
\begin{center}
\includegraphics[width=6.0cm,angle=0]{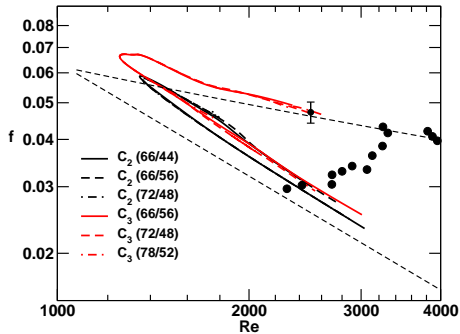}
\end{center}
\caption[]{Coherent structures in a friction factor vs. Reynolds number plane for pipe flow.
$C_2$ and $C_3$ indicate states with two and three vortex pairs, respectively, and the
numbers in parenthesis indicate different resolutions (see \cite{Fai03}). The dots indicate
observations on a turbulent flow, and the upper and lower dashed lines are extrapolations
of the turbulent and laminar behaviour, respectively.
\label{bif_diagram}
}
\end{figure}

In the case of pipe flow new persistent flow structures besides 
the laminar profile have indeed been identified \cite{Fai03,TW_bristol}. 
They are of the type of full three-dimensional travelling waves, 
see Fig.~\ref{coherent_states}. Three-dimensionality is required,
since velocity fields with translational symmetry in the
downstream direction decay and cannot sustain any non-trivial flow
state. The first structures identified were highly symmetric,
and contained several pairs of vortices. Among these, the structure
with the lowest critical Reynolds number is the threefold set of vortices,
with a bifurcation near $Re=1250$. It is followed by the bifurcation
of the twofold state at Reynolds number $Re=1350$, and more at higher
Reynolds number. Interestingly, the friction factors are sometimes
higher than the turbulent ones, but they tend to approach the laminar values
for higher Reynolds number, Fig.~\ref{bif_diagram}. Less symmetric 
structures may also be identified, and some of them extend to lower
Reynolds numbers \cite{Pri07}.
 
All these states actually belong to continuous families, since
they tolerate some stretching and compression for Reynolds number above
their point of bifurcation, \cite{Fai03,aberdeen}. However, their optimal 
wavelengths at the bifurcation of $2.5R$ and $4.2R$ for three and two 
pairs of vortices, respectively, \cite{Fai03}, can fit well into the
short domains studied in the next section.

The bifurcation scenario in pipe flow, however, differs from the usual one in that 
both states that appear in the bifurcation are immediately unstable \cite{aberdeen}: 
it is like a saddle node bifurcation in an unstable subspace.  This has two consequences.
As a first one we note that the absence of stable patterns and the hyperbolic
dynamics on the saddle naturally explain the chaotic dynamics.
Nevertheless, the coherent structures just discussed can show 
up transiently in the dynamics \cite{science},
where they can be detected by application of the indicator (\ref{indicator}).
Studies on time series then show that about 20 per cent of the time the
flow is dominated by the presence of these structures \cite{Sch07b}.

Another consequence of this deviation from the standard saddle-node 
transition scenario
is the fact that the stable manifold
of the saddle is not of co-dimension one and therefore does not subdivide state space into 
the domains of attraction of the laminar and the turbulent dynamics. 
Moreover, the saddle does not collide with the laminar profile for 
any finite Reynolds number: the laminar profile remains stable
against infinitesimal perturbations for all Reynolds numbers \cite{Mes03}.
This raises interesting questions about the boundary between laminar and turbulent
dynamics to which we turn in section \ref{sec:edge}, 
and it opens up the possibility that the turbulent
dynamics is not persistent but transient, a feature we analyze in the next section.

\section{Lifetimes}
\begin{figure}
\begin{center}
\includegraphics[width=5.cm,angle=0]{L10PDF_newscale_withleg.eps}
\includegraphics[width=5.cm,angle=0]{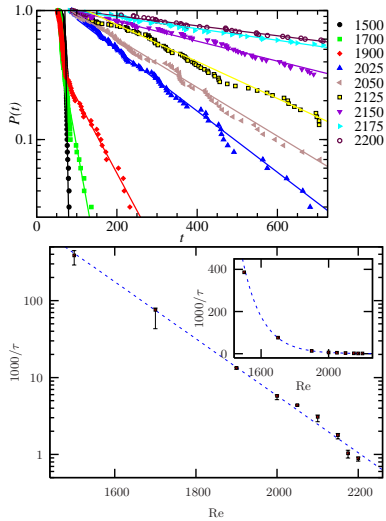}
\end{center}
\caption[]{Lifetime distributions for pipe flow, and variation of mean lifetime with $Re$
for a pipe segment of length $10R$. Top panel: Probability to still be turbulent after 
some time $t$ for several Reynolds numbers in a semi-logarithmic plot. The straight lines 
indicate exponential fits to the tails of the distributions. Bottom panel: Inverse 
characteristic lifetimes $\tau$ extracted from the exponential fits shown in the top panel. 
Times in this plot are measured in units of $2R/U$  
with $R$ the pipe radius and $U$ 
the mean downstream velocity.
\label{lifetimes}
}
\end{figure}

In the usual subcritical bifurcation scenario the stable node forms
the basis for an attracting subset in state space, where chaos can set
in as a result of further bifurcations.  However, in the absence of a
stable object, i.e. if the bifurcation happens within a subspace
that is already unstable with respect to perpendicular deviations, the
persistence of the dynamics in the turbulent region has to be
investigated separately. Indeed, studies by Brosa \cite{brosa89} first
suggested the intriguing possibility that turbulence might be transient.
As a first step in this direction we study
lifetimes of a sample of initial conditions in the turbulent region.
Numerical \cite{Fai04,Hof06,Sch07a} and
experimental studies \cite{Mul05,Mul06,Pei06,Pei07,Hof06} 
show that in this region the lifetimes are exponentially distributed.  
Asymptotically, for long times, the probability to find a turbulent state 
after a time $t$ is given by 
\beq
P(t)\propto e^{-(t-t_0)/\tau}\qquad \mbox{for} \qquad t>t_0\;, 
\eeq 
with a characteristic decay time $\tau$.  The
exponential distribution for $P(t)$ implies that the probability density 
$p(t)=$ $d\ln P(t)/dt$ to decay at a time $t$ is constant: the flow has no
memory and decays at unpredictable moments in time. This constant
decay probability is strong evidence for the
formation of a chaotic saddle that supports transient turbulent
dynamics. The characteristic decay time $\tau$ increases rapidly with Reynolds
number, but it does not seem to diverge at any finite $Re$.  This is
confirmed in experiments on long pipes and in numerical simulations as 
shown in Fig.~\ref{lifetimes} \cite{Hof06}.
Other experimental studies on shorter pipes \cite{Mul05,Mul06,Pei06,Pei07} and
simulations for five Reynolds numbers \cite{Wil07} have been
interpreted using a variation $\tau(Re)\sim a/(Re_c-Re)$: however, the
extracted parameter $a$ and the critical Reynolds number $Re_c$ vary
over a wide range, leaving questions whether the asymptotic regime has
indeed been reached. Moreover, a reanalysis of the data \cite{arxiv}
shows that the numerical simulations are fairly close to the
experimental results from Hof et al \cite{Hof06}.  The comparison
between the data suggests the interesting possibility that 
a global bifurcation near Reynolds numbers of about 1850 
(see also \cite{Ben07})
might result in a quantitative change of the scaling function
parameters. Perhaps previously separated parts of the turbulence
supporting scaffold become connected to raise the lifetimes, but without
actually turning the turbulent saddle into an attractor.

The observation of transient turbulence complicates the identification of 
a critical Reynolds number for the transition to turbulence: ideally, this
would be the Reynolds number where the flow becomes persistently turbulence. 
The absence of a divergence in the lifetimes rules this definition out. 
Alternatively, one might require that more than half of all initial conditions
remain turbulent up to some time: then the critical Reynolds number depends
on the probability and the observation time. Fortunately, these dependences
are fairly weak on account of the rapid increase with Reynolds number: 
with the data from \cite{Hof06}, the characteristic time $\tau$ (measured
in units of $2R/U$) reaches
1000 for a Reynolds number of 1924, and 2000 already 
for 1944.

\section{Edge of chaos}
\label{sec:edge}

\begin{figure}
\begin{center}
\includegraphics[width=5.cm,angle=0]{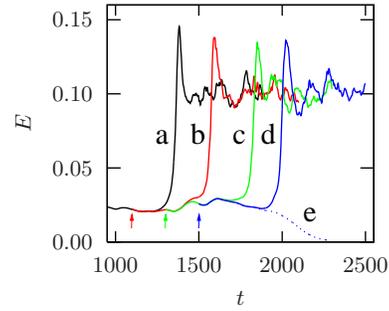}
\end{center}
\caption[]{Visualization of the edge state tracking algorithm for pipe flow. 
  The edge state is bracketed by a pair of initial conditions in which 
  one initial condition decays and the other becomes fully turbulent.  
  After about $200$ time units the approximation is refined
  by choosing a new pair of approximating trajectories. The temporal modulations
  in energy indicate that the edge state maintains some dynamics and does not 
  become a travelling wave.
\label{tracking}
}
\end{figure}

\begin{figure}
\begin{center}
\includegraphics[width=3.cm,angle=0]{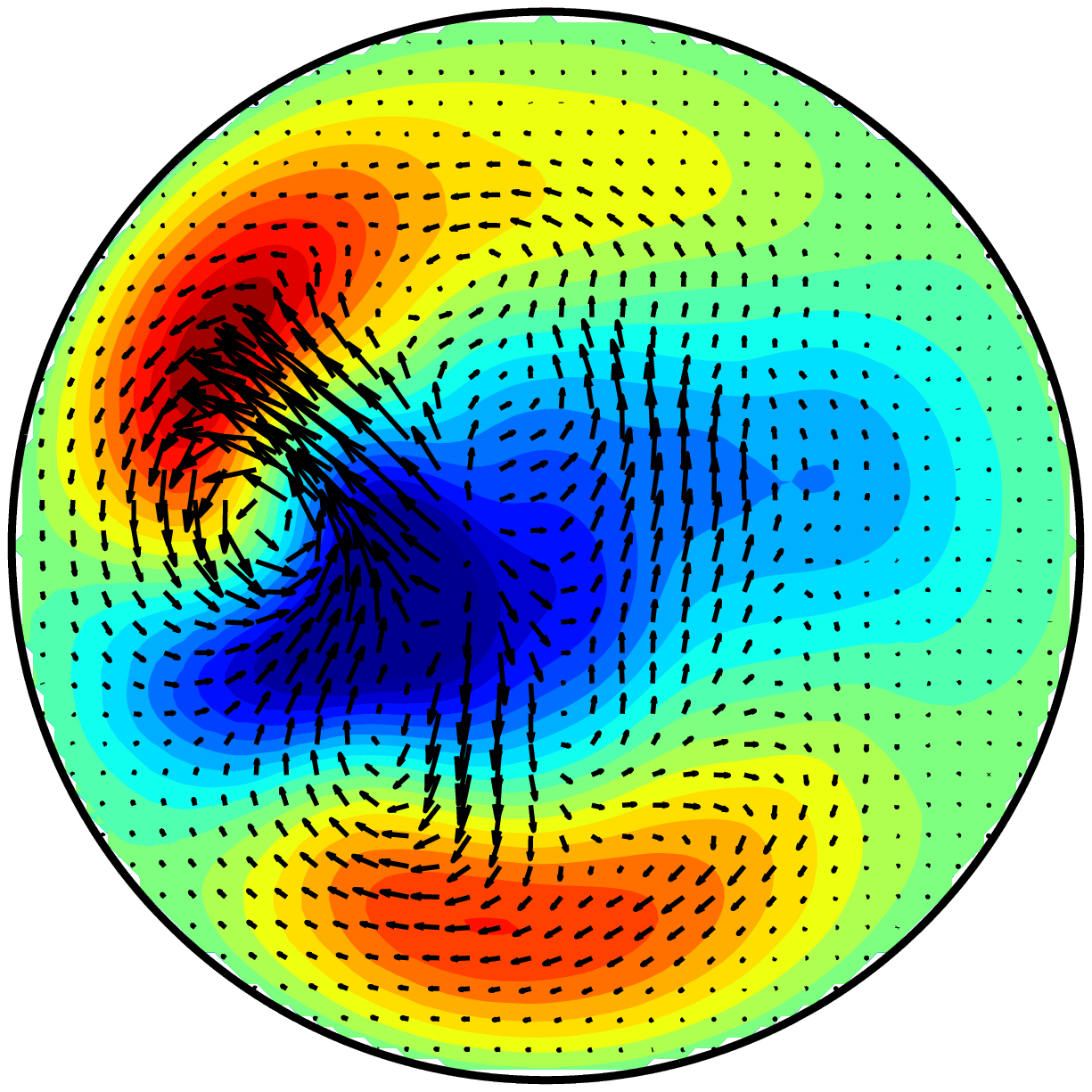}\qquad
\includegraphics[width=3.cm,angle=0]{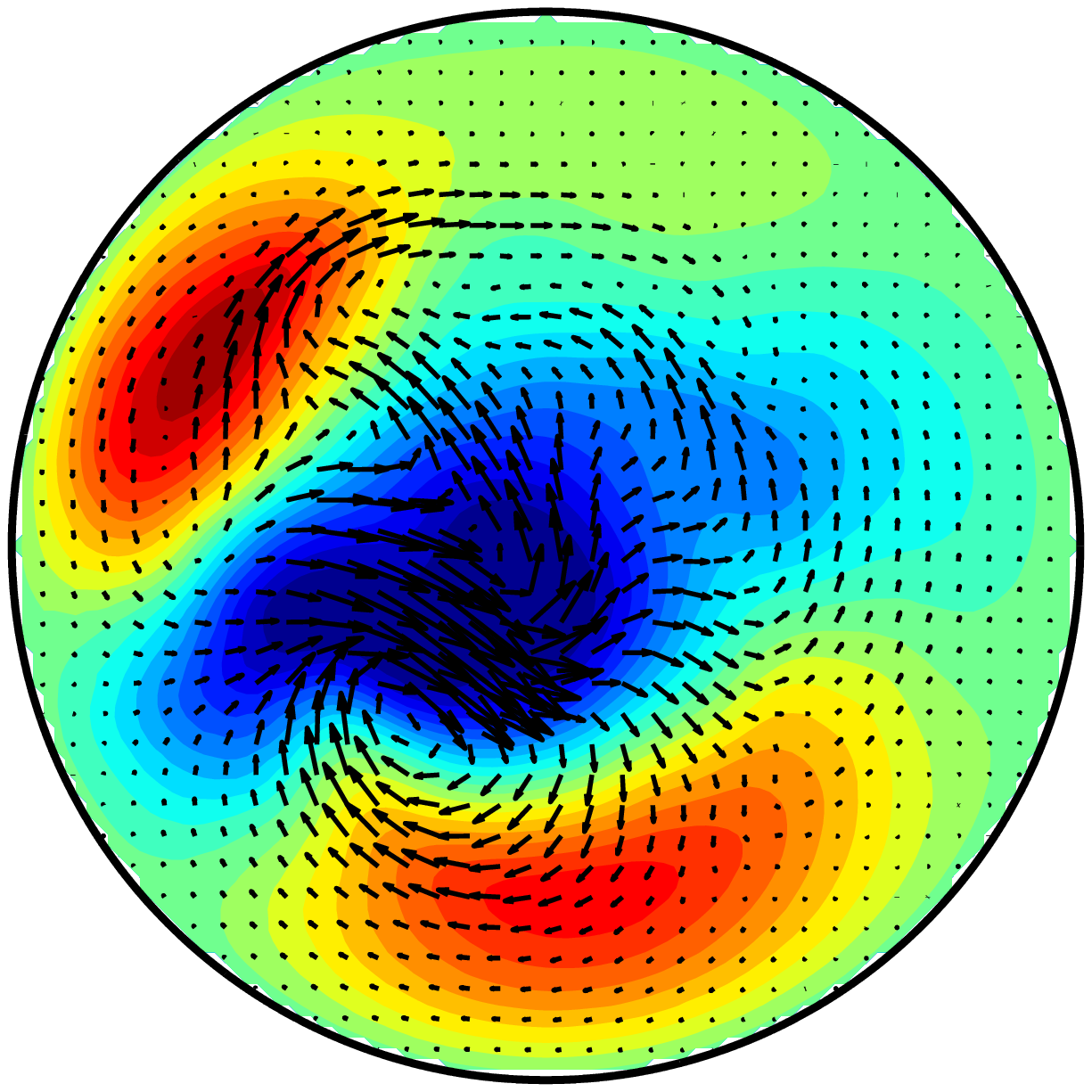}
\end{center}
\caption[]{Two snapshots of the edge state at $Re=2160$. The snapshots clearly show
the oscillating motion of the center vortices between the two high-speed
regions near the walls.
\label{edge_states}
}
\end{figure}

The coexistence of a chaotic saddle carrying the turbulent dynamics and
the linearly stable fixed point surounded by its basin of attraction
naturally suggests to look for the boundary between the two objects. 
In the case of coexisting attractors, this would be the basin boundary.
However, in order to cover the case of a transient dynamics, 
the concept of the \emph{edge of chaos} 
was introduced \cite{Sku06,schneider_06,Sch07a,Vol08}: 
when crossing
from the laminar to the turbulent side by increasing the amplitude of the
perturbation one notes and smooth increase in lifetimes on the laminar side
and a sensitive dependence on initial conditions on the turbulent side \cite{Fai04}. 
Inbetween is a first point where the lifetime becomes infinite. Because of
the chaotic variations on the turbulent side, this point sits on the 
edge of chaos. The location of this edge of chaos in state space can
be expected to be fairly complex: folds have been identified \cite{Sch07a}
and fractal properties cannot be ruled out \cite{Sch97,Moe04,Vol08}.
Interestingly, different points on the edge of chaos can
be dynamically connected, and it makes sense to probe for the dynamics on 
the edge of chaos. The evolution of the system then shows that there is 
a relative attractor: it is attracting for initial conditions in the edge,
but globally unstable because of the tendency to either swing up to the
turbulent dynamics or to return to the laminar one. We call this
relative attractor the \emph{edge state}, as it is embedded in the edge of 
chaos. As mentioned before, it extends the concept of basin boundary
for attractors to the situation with an attractor and a saddle,
see \cite{Vol08} for further discussion. The boundary and its dynamics
is also relevant for determining the smallest perturbation sufficient
to trigger turbulence, e.g. \cite{alvaro},

On the practical side, the boundary can be tracked by straddling it with
a pair of trajectories where one returns to the laminar state directly,
and the other swings up to the turbulent state. Since the dynamics
in the edge is unstable, the two trajectories will separate. If the
separation becomes large enough, a new pair of trajectories can be
found by refinement, see Fig.~\ref{tracking}. 
This way the pair will always stay close to
the edge and will approximate a trajectory the never leaves it.
The velocity fields for these trajectories become
progressively simpler as time goes on \cite{schneider_06,Sch07a}. 
In the end, a state consisting of two high speed streaks,
a low speed streak in the middle, and a pair of vortices that drives the
streaks, is obtained, see Fig.~\ref{edge_states}. Interestingly, the two vortices in the
middle are dynamically active, but do not leave this region. This type
of state can be followed up to higher Reynolds numbers.

\section{Conclusions}
The studies on the transition to turbulence in pipe flow have confirmed
a few expectations (like a subcritical transition scenario), but have also
revealed unexpected properties: the lack of stability of the coherent structures,
and their transient appearance in turbulent flows, the non-persistent nature
of the turbulent state, the complicated boundary between laminar
and turbulent, etc. The effort invested into understanding this transition
can also be expected to be helpful in other situations: plane Couette flow
is perhaps the next flow that comes to mind, as it shares with pipe flow
the linear stability of the laminar profile to infinite Reynolds numbers.
Pressure driven plane Poiseuille flow does have a linear instability,
but at a Reynolds number above the one where turbulence appears. 
All these flows share many similarities, and can be profitably studied
with the concepts presented here.\\

Financial support by the Deutsche Forschungsgemeinschaft is
gratefully acknowledged.
%
%


\end{document}